\documentclass[12pt,a4paper]{article}

\usepackage{times}
\usepackage{graphicx}
\usepackage{euscript,amssymb}
\usepackage{amsfonts}
\usepackage{amssymb,amsmath}
\usepackage{mathrsfs}
\usepackage{fancybox}
\usepackage[latin1]{inputenc}
\usepackage{pst-all}
\usepackage{epsfig}
\usepackage{rotating}
\usepackage{subfig}
\usepackage[T1]{fontenc}
 
\newcommand{\be}{\begin{equation}}
\newcommand{\ee}{\end{equation}}
\newcommand{\bea}{\begin{eqnarray}}
\newcommand{\eea}{\end{eqnarray}}

\newcommand{\bdm}{\begin{displaymath}}
\newcommand{\edm}{\end{displaymath}}

\begin{document}


\begin{titlepage}

  \noindent 
\begin{center}
  
  {\Large \bf G\"odel's undecidability theorems}
  \vskip 2mm
  {\Large \bf and the search for a theory of
    everything}\footnote{Slightly elaborated version of a Prize
    winning essay awarded by the Kurt 
    G\"odel Circle of Friends Berlin with the support of the
    University of Wuppertal, first published in
    https://kurtgoedel.de/kurt-goedel-award-2023/} 

  \vskip 1cm

{\bf Claus Kiefer} 
\vskip 0.4cm
University of Cologne, Faculty of Mathematics and Natural Sciences,\\
Institute for Theoretical Physics, Cologne, Germany
\vspace{1cm}

\end{center}

\begin{abstract}
I investigate the question whether G\"odel's undecidability theorems
play a crucial role in the search for a unified theory of physics. I
conclude that unless the structure of space-time is fundamentally
discrete we can never decide whether a given theory is the final one
or not. This is relevant for both canonical quantum gravity and string theory. 
  \end{abstract}

\end{titlepage}

\vskip 4cm

\noindent Was mich urspr\"unglich interessiert hat, ist die
Erkl\"arung der Erscheinungen des Alltagslebens aus h\"oheren
Begriffen und allgemeinen Gesetzm\"a\ss igkeiten, daher
Physik. (\textsc{G\"odel}~2020, p.~81)

\vskip 5mm

\noindent English translation: I have been originally interested in
explaining the 
  phenomena of everyday life in terms of higher concepts and general
  regularities, hence physics. (\textsc{G\"odel}~2020, p.~347)

  \newpage

\section{The search for unification}

In his inaugural lecture for the Lucasian Chair of Mathematics at the University of
Cambridge, the eminent theoretical physicist Stephen Hawking expressed the
following vision for the future (\textsc{Hawking}~1980):
\begin{quote}
In this lecture I want to discuss the possibility that the goal of
theoretical physics might be achieved in the not too distant future,
say, by the end of the century. By this I mean that we might have a
complete, consistent, and unified theory of the physical interactions
which would describe all possible observations.
\end{quote}
This was in 1979. In retrospect we can say that such a unified
theory of the physical interactions was not available in the year
2000, nor is it available today. The ``dreams of a final theory'', in
the words of Steven Weinberg 
(\textsc{Weinberg}~1993), have not yet been materialized. In more sober
words, the reductionist programme of physics has not yet come to an
end. Whether it will ever come to an end, is an open issue
and is the topic of this essay. As we shall see, G\"odel's
undecidability theorems will play a crucial role in this investigation.

In view of the history of physics, the reductionist programme seems
natural and straightforward. What have earlier been separate theories
(and models) were later recognized as special cases of a common
theory: electricity, magnetism, and geometric optics, for example,
were recovered as particular limits of the theory of electrodynamics,
developed in the 19th century by James Clerk Maxwell. All known
effects in these areas could be deduced from one fundamental set of
partial differential equations -- Maxwell's equations. Another example
is the partial unification of the electromagnetic and weak
interactions to the electroweak interaction that together with the
strong interaction forms what is today called the {\em Standard Model} of
particle physics.

The Standard Model was constructed at the end of the
1960s and the beginning of the 
1970s by Weinberg and others. It is in this context that
Hawking's speech must be seen. As part of the unification
attempts in the 1970s, models of {\em supergravity} were constructed,
which aim at a unification of gravity -- so far successfully described
by Einstein's theory of general relativity (GR) -- with the other
interactions. The {\em super} in this word refers to a hypothetical
symmetry between fermions (to which electrons and protons belong) and
bosons (to which photons and gravitons belong) called
supersymmetry. Hawking, in 1979, speculated that the final theory has
the form of a supergravity theory. Later, after it had become clear
that this hope has remained unfulfilled, he favoured string theory and
M-theory (a certain extension of string theory).\footnote{In a talk
  given in 2002, however, he speculated that we shall never find an
  ultimate theory, and even mentioned G\"odel's theorem(s) as a reason
  for that (\textsc{Hawking}~(2002))}.
  
  String theory became
popular in 1984 when indications were found that properties of the
Standard Model   
are present in this theory. So far, however, a recovery of the
Standard Model from string theory remains an unfulfilled dream. String
(and M-) theory contain supersymmetry as a necessary 
ingredient to ensure its consistency, but no sign of supersymmetry was
found to date in 
experiments performed at the Large Hadron Collider (LHC) at Cern,
Geneva, and elsewhere.  

When we speculate about a unified theory, we implicitly assume that we
deal with a final theory, that is, we assume that there is no deeper
structure of physical theories. In Weinberg's words, a final theory is
characterized as follows:

\begin{quote}
  A final theory will be final in only one sense--that it will bring
  to an end a certain sort of science, the ancient search for those
  principles that cannot be explained in terms of deeper principles.
  (\textsc{Weinberg}~1993, p.~13)
  \end{quote}

One may add here that a truly final theory should also be rigid in the
sense that small changes in its parameters do not change its essential
structure. But can we really decide whether a given unified theory is
final in this sense?

Physical theories are formulated in the language of mathematics, so
the question of unification in physics is deeply related with the
construction of a `unified' mathematical language. At the beginning of
the last century, it was the mathematician David Hilbert who
attempted to construct a unified mathematical language. He searched
for an axiomatic foundation of geometry and finally the whole of
mathematics. Hilbert belonged to what is today called the formalistic
school. There, axioms are no longer `obvious' statements in the sense
of Euclid but arbitrary formal settings whose justification lies in
the success of finding a unified scheme. In view of Ludwig
Wittgenstein's much later philosophical investigations, one may call
Hilbert's formal system a {\em Sprachspiel} (language-game). In mathematics, such a
language-game is constrained by two important properties that will
play an important role below: completeness and
consistency. Completeness means that every statement that can be
formulated in this formal system can be proved or disproved;
consistency means that there are no logical contradictions between
different statements in the formal system. 

Hilbert's dream of unification was not restricted to mathematics. He
intended to generalize this to physics by providing a unified
mathematical language for the physical interactions. The only known
fundamental interactions at the time were gravity and
electrodynamics. The latter was described by Maxwell's equations, but
what about the former? Here, the big achievement was Einstein's theory
of general relativity (GR), completed in November 1915. Its central
equations, Einstein's field equations, so far describe all known
gravitational phenomena (or are at least not obviously in conflict with
them). Hilbert arrived at those field equations around the same time
by postulating a geometric variational principle, starting from what
today is known as the Einstein--Hilbert action. But in contrast to
Einstein, he envisaged to derive all field equations of physics by
such a geometric variational principle. This is why his publication
bears, not very modestly, the title ``The Foundations of Physics''
(\textsc{Hilbert}~1915).

The goal of unifying gravity with electrodynamics by a geometric
theory in the spirit of GR was not achieved by
Hilbert. Nor was it achieved by Einstein, who spent most of his later
years with attemps of finding a unified field theory. In retrospect,
we can claim mainly two reasons for this failure. 
First, neither Hilbert nor Einstein took into account
the microscopic interactions known as weak and strong
interactions, which were studied from the beginning of the 1930s onward. And
second, perhaps even more important, quantum theory was not
addressed in these attempts, a theory
not known at the time of Hilbert's 1915 article, but
known and experimentally established at the time of Einstein's later
work in the 1940s and 1950s. Connected with this second point is the
question of the space-time continuum, used by Einstein in the
traditional sense, and its fate in a unified theory encompassing
quantum theory. We shall come back to this point below.  

Hilbert was very optimistic towards the materializability of his axiomatic
programme. In this, he was an antagonist of the physician and
physiologist Emil du Bois-Reymond
who had formulated his famous
{\em Ignoramus et ignorabimus} (we do not know and will not know) in his
1872 keynote address {\em \"Uber die Grenzen des Naturerkennens} (On the limits
of science). Du Bois-Reymond was convinced that there were fundamental
limits to our knowledge of Nature and natural laws. Already in 1900,
at a major conference on mathematics in Paris,\footnote{It was at
  that conference where Hilbert presented his list of 23 important
  unsolved problems in mathematics.} Hilbert emphasized that
in his opinion there is no {\em ignorabimus} in mathematics.
Thirty years later, just one year before the publication of G\"odel's
undecidability theorems, he emphasized his standpoint again in
strong words in a radio address:\footnote{See and ``hear''
  www.maa.org/press/periodicals/convergence/david-hilberts-radio-address,\\
  where also the German transcription and the English translation can
  be found.}
\begin{quote}
  We must not believe those, who today with philosophical bearing
  and a tone of superiority prophesy the downfall of culture and
  accept
  the {\em ignorabimus}. For us there is no {\em ignorabimus}, and in
    my opinion even none whatever in natural science. In place of the
    foolish {\em ignorabimus} let stand our slogan: We must know, We
    will know.\footnote{German original: ``Wir dürfen nicht denen
      glauben, die heute mit philosophischer Miene und überlegenem
      Tone den Kulturuntergang prophezeien und sich in dem Ignorabimus
      gefallen. Für uns gibt es kein Ignorabimus, und meiner Meinung
      nach auch für die Naturwissenschaft überhaupt nicht. Statt des
      törichten Ignorabimus heiße im Gegenteil unsere Losung: Wir
      müssen wissen, Wir werden wissen.'' This slogan is engraved on
      his tombstone in G\"ottingen.}
    \end{quote}

So will we one day know the final theory of physics or will there be a
final {\em ignorabimus}? In modern days, especially in particle
physics, a final theory is sometimes referred to as a ``theory of
everything'' (TOE). This wording implies that such a theory will not
only provide a unified theory of physics, but -- at least in principle
-- a theory for all possible effects in chemistry, biology, and maybe
even beyond. The crucial word here is {\em in principle}. As we know
today, even within physics we cannot, in general derive effects at an
effective level from a fundamental theory. For example, nuclear
physics is thought to arise as a limit from quantum chromodynamics
(QCD), our fundamental theory of the strong interactions. But in
practice, the formalism is so complicated that this can hardly be
done; this is why models such as the nuclear shell model are still
essentially used in the everyday work of a nuclear physicist. It is
thus evident that this limitation holds even stronger for
biology. There, we deal with even more complex systems, and concepts
such as {\em synthetic biology} are more powerful than biological laws
arising from fundamental physics (\textsc{Weitze} and
\textsc{P\"uhler}~2014). Nevertheless, the important question for us is
whether a final theory exists in principle, independent of these
practical limitations. To address this question, it is necessary to
clarify the relation between mathematics and physics, which is subject
of the next section.

\section{Mathematics and physics}

In a well-known article, the Nobel Prize winner Eugene Wigner
speculates about the, according to his opinion, unreasonable effectiveness of
mathematics in the natural sciences, in particular physics
(\textsc{Wigner}~1960). Why is it that physical phenomena can be
described by mathematical equations? And why exists, it seems, a
small set of equations, such as Einstein's equations, Maxwell's
equations, and the Schr\"odinger equation, that lie at the foundation
of this description? Already Galileo, in his {\em Il Saggiatore},
envisaged the Universe as being written in mathematical language,
which for him was the language of triangles, circles, and other
geometric figures. Our modern mathematical description of physics dates back to
Isaac Newton, the second Lucasian Professor of Mathematics at the
University of Cambridge. Wigner, in his Nobel Prize speech of 1963,
emphasized the surprising discovery of Newton's age that the laws of
Nature can be clearly separated in dynamical laws and initial
conditions. The dynamical laws are given by differential equations up
to second order in space and time. They thus leave room for initial
(or, more generally, boundary) conditions, which are not fixed by the laws and which
thus represent contingent features of our Universe.

But why is mathematics so effective? A full answer is elusive, but a
partial answer may lie in the role symmetries play at a fundamental
level. The structure of the Standard Model is given and, in fact,
strongly constrained by gauge invariance. This is an internal
symmetry that acts on all the quantum fields representing particles,
thus connecting them in a non-trivial manner. Gravity is not contained
in the Standard Model. It is described by Einstein's field equations
which displays another type of symmetry (or, more properly,
invariance) -- the diffeomorphism invariance of space-time. Hereby
is meant the mathematical exact expression for coordinate invariance:
space-time points have no meaning independent of the dynamic degrees
of freedom representating geometry and matter fields.

There are quite a few theoretical physicists who regard the fundamental
mathematical equations as `beautiful'. This sense of beauty is
connected with the symmetries or invariances that these equations
exhibit. It only applies to the equations themselves; their solutions
as well as approximations following from them may be lengthy,
complicated, and ugly. In the words of \textsc{Weinberg} (1993),
p.~131: ``It is when we study truly fundamental problems that we
expect to find beautiful solutions.'' Paul Dirac, another holder of
the Lucasian chair of Mathematics, who invented the equation named
after him, went even further in claiming that beauty in one's
equations is more important than compatibility with experiment. Most
physicists would not support such a strong view because it brings the
danger of formulating equations devoid of empirical content. In 1931, Dirac
himself had presented an even more symmetrical version of Maxwell's
equations, which in addition to the usual electric charges contains
magnetic monopoles. But such monopoles have never been seen and they may,
in fact, not exist at all. Still, the question remains what is the
structure of the fundamental equations of a
unified and final theory. String or M-theory, in its present state,
does not exhibit `beautiful' equations nor is it based on an
aesthetically appealing fundamental principle. But can we say
something about the possible mathematical structure of a final theory?
The mathematician Kurt G\"odel, too, expressed this idea of beauty:

\begin{quote}
  The beauty in the presentation of a subject lies in first
giving general (abstract) concepts and possibly a theory and \underline{then}
the application to the empirical \ldots For
the same reason, the beauty in physics is the explanation of everyday
phenomena. Hence also the name ``knowledge''.
(\textsc{Gödel} (2021), p.~229)\footnote{German original:
   Das Schöne an der Darstellung einer Sache ist, zunächst
allgemeine Begriffe (abstrakte) und eventuell ihre Theorie
zu geben und \underline{dann} die Anwendung auf die {\em Empirie}. \ldots
Dasselbe ist der Grund dafür,
dass das Schöne an der Physik die Erklärung der alltäglichen
Erscheinungen ist. Daher auch der Name {\small $\gg$}Erkennen{\small $\ll$}.
(\textsc{Gödel} (2021), p.~70)}
\end{quote}

It is most likely that also Hilbert in his search for unification in both
mathematics and physics was driven by some concept of beauty. But
Hilbert's programme received a severe blow when G\"odel presented his
undecidability theorems (\textsc{G\"odel 1931})\footnote{For an
  English translation, see \textsc{G\"odel} 
  (1986), pp.~145~ff.}. This blow applied to mathematics; whether it
also applies to physics and to the dreams of a final theory is the
subject of this essay.
In the words of Douglas Hofstadter, 
G\"odel's first undecidability theorem can be paraphrased as follows:

\begin{quote}
  All consistent axiomatic formulations of number theory include
  undecidable propositions. (\textsc{Hofstadter} 1982, p.~17)
\end{quote}

Hofstadter compares this theorem with a pearl that is buried in an
oyster, the oyster standing for the mathematical proof of this theorem,
which makes essential use of self-referring statements. The theorem
has the far-reaching consequence that in any sufficiently complex
axiomatic system (complex enough to contain the arithmetics of natural
numbers) there are statements that can neither be proved nor
disproved. 
So, as Hofstadter continues to write,  ``provability is a {\em weaker}
notion than truth, no matter what axiomatic system is involved.''
Traditionally, the assumption entertained by mathematicians always was
that a certain statement within a formal system can either be proved
or its negation can be proved; now there is a third option 
called {\em undecidable}. 

At the end of his article, G\"odel announces what today is called his
Second Incompleteness Theorem, which can be paraphrased as ``if a
sufficiently complex axiomatic system containing the arithmetics of
natural numbers is consistent (free of contradictions), it is
impossible to prove this consistency within the system itself''. Thus,
Hilbert's ambitious programme of developing a complete and consistent
formal scheme for the whole of mathematics cannot be accomplished. 

An important application is the halting
problem, which is the problem of whether there exists a general algorithm
with which one can decide whether an arbitrary 
programme with arbitrary input will finish after a finite number of steps. This
was shown by Alan Turing in 1936 to be undecidable.  Apart from the
halting problem,
the perhaps most important example for an undecidable problem is the continuum hypothesis
(CH). It goes back to the mathematician Georg Cantor and can be
phrased as stating that there is no cardinal number between the set of
natural numbers $\mathbb N$ and the set of real numbers $\mathbb
R$. Assuming the validity of the axiom of choice, it can also be
stated in the form that the cardinality of the power set
$2^{\aleph_0}$ is equal to $\aleph_1$, where $\aleph_0$ is the
cardinality of the natural numbers and $\aleph_1$ the cardinality of
the real numbers. The continuum hypothesis is actually the first entry in
Hilbert's famous list of 23 problems presented in 1900.
It is a statement about
numbers; the term continuum comes from the idea, mostly taken as
granted, to identifying $\mathbb 
R$ with the points on a line. That such an identification can be
questioned is a later insight and will be discussed below in connection
with its relevance for unified theories. 

The continuum hypothesis was shown to be undecidable by Paul Cohen in
1963. He could prove that it cannot be proved or disproved from the standard
axioms of set theory (the Zermelo--Fraenkel axioms). G\"odel, who
subscribed philosophically to what one may call platonic realism,
believed that the CH is true or false, even if no proof or disproof
can be given within standard set theory.\footnote{See his
article ``What is Cantor's continuum problem?'', first published in 1947,
see \textsc{G\"odel} 1990, pp.~176--187,
and later published in a revised version in 1964, see \textsc{G\"odel}
1990, pp.~254--270.} He envisaged that there will be a future powerful
axiomatic system within which a decision can be made.

The story of G\"odel's theorems and their consequences for the history
of mathematics has been told many times, and their is no point to
repeating it here.\footnote{See, for example, \textsc{Hofstadter}
  (1982) or the detailed editorial
  notes in \textsc{G\"odel} (1986) and \textsc{G\"odel} (1990).}
We are interested, instead, in discussing the relevance of 
G\"odel's results for the construction of a unified physical theory
(e.g. string theory), a story that so far is unduely
neglected.\footnote{As Roger Penrose has remarked: ``It is my own
  personal opinion that we shall find that computability issues will
  eventually be found to have a deep relevance to future physical
  theory, but only very little use of these ideas has so far been made
  in mathematical physics.'' (\textsc{Penrose} 2004, p.~378). This
  essay is an attempt to fill this gap.}

In the last section below, we will investigate the relevance of the continuum
hypothesis for mathematical models of space-time. Before
doing so, we have to understand the role of quantum theory in the search
for a unified theory, something not yet attempted by Hilbert and
Einstein in their searches for a unified theory of physics.

\section{The role of quantum theory}

The non-relativistic version of quantum theory was constructed ten
years after Einstein's (and Hilbert's) work on general relativity, in
the years 1925--27. Generalizations to field theory were started later and are not yet
fully completed. The reason for this unfinished state of affairs lies
in the infinitely many degrees of freedom of quantum field theory:
sophisticated mathematical schemes of regularization and
renormalization were invented to deal successfully with formally
infinite expressions, but the problem of the infinitely small (and,
thus, the continuum) remain unsolved. The Standard Model of particle
physics mentioned above is such a quantum field theory.

Mathematically, the Standard Model is a gauge theory (Yang--Mills
theory) and contains in particular the theory of strong interactions
(QCD). An important issue is whether one can prove within this 
theory the observed confinement of quarks. Interestingly, this problem
seems to be undecidable (\textsc{Cubitt} {\em et al.} 2015). In fact,
this problem belongs to the general class of spectral-gap problems,
under which one understands the question whether there is a gap
between the ground state energy of a given system and its first excited state
or not. \textsc{Cubitt} {\em et al.} (2015) were able to relate this
problem to Turing's halting problem, from which the undecidability of
the spectral-gap problem follows.\footnote{A concrete construction of
  a Hamiltonian operator whose spectral gap is undecidable is given in
  \textsc{Cubitt} (2021).} In this way, G\"odel's theorems enter the
Standard Model of particle physics.

In their proof, \textsc{Cubitt} {\em et al.} (2015) make essential use
of the thermodynamic limit, that is, the limit where the number of
degrees of freedom tends to infinity. This limit was taken because
of its relevance for quantum phase transitions: the transition from a
gapless to a gapped situation (or {\em vice versa}) can occur at
arbitrary large (and uncomputable) values for the parameter describing
the thermodynamic limit.

An application to field theory was recently presented by
\textsc{Tachikawa}~(2023) in the context of supersymmetry. His line of
arguments is interesting. He poses the question whether one can prove
that a certain 
supersymmetric model (Wess--Zumino model) can entail supersymmetry
breaking. He related this problem to Hilbert's tenth problem about
Diophantine equations, which is known to be undecidable, and concluded
that his question about supersymmetry breaking is undecidable, too. 

The notion of infinity thus plays an essential role in all these
considerations. It seems that so far in physics only the cardinality
$\aleph_1$ of the real numbers play a role,\footnote{The cardinality
  of the complex numbers $\mathbb C$, which play a major role in
  quantum theory, is the same as the cardinality of $\mathbb R$.} but this is already
sufficient to give rise to the problems discussed here. Early on,
\textsc{Komar} (1964) remarked, on the basis of G\"odel's theorem but
without going into technical details, 
that the question of whether two states in quantum field theory are
macroscopically distinguishable or not, is undecidable. His arguments
only work for systems with infinitely many degrees of freedom, thus
assuming a space-time continuum.

These issues are connected with another most important problem in
quantum theory: the problem of the classical limit. A central (and
perhaps its most characteristic) feature of this theory is the
superposition principle -- the sum of two physically allowed quantum
states is again an allowed state. This immediately leads to the
occurrence of weird macroscopic states such as Schr\"odinger's
cat. The fact that such states are not observed was a perennial puzzle
of the theory. One way towards its solution is the assumption of a
major modification of quantum theory in the form of a wave-function
collapse. Collapse models are studied in detail (\textsc{Bassi} {\em
  et al.} 2013), but so far none has been experimentally
established. Another way is the realistic modelling of the system's
environment which can lead to the formation of correlations rendering
the superposition unobservable by transfering the information about it
in entanglement between system and environment. This process is called
decoherence and is experimentally well established (\textsc{Joos} {\em
  et al.} 2003).

Decoherence is also of relevance for an interesting discussion about
the origin of consciousness. Roger Penrose, in collaboration with the
anesthesiologist Stuart Hameroff, developed the idea that quantum
superpositions in neural microtubules in the brain are responsible
for the emergence of consciousness; this was again concluded by making a
connection with the undecidability of the halting problem
(\textsc{Penrose} 1994). The brain,
according to Penrose and Hameroff, works in a non-algorithmic way and can thus
give rise to free will.\footnote{Incidentally,
  the question of free will is also included in the seven ``world riddles''
  formulated by du Bois-Reymond.}

It is of some interest to note that other scientists had speculated
earlier about a connection between consciousness and quantum
theory. The mathematician John von Neumann as well as the above
mentioned Eugene Wigner entertained the idea that consciouness is, in
fact, responsible for the occurrence of a wave-function collapse,
avoiding in this way paradoxical states such as Schr\"odinger's
cat. Wigner gave up this idea in the 1970s after the process of
decoherence was discovered by the physicist Dieter Zeh.\footnote{See,
  for example, \textsc{Kiefer} (2022) for a discussion.} 

Decoherence did not only lead Wigner to change his mind, but also to
render the Penrose--Hameroff scenario unlikely. As Max Tegmark has
shown, the decoherence times for possible quantum superpositions in
the brain are much shorter than standard time-scales used for
conscious processes, thus leading to their irrelevance 
(\textsc{Tegmark}~2014).

The question of the quantum-to-classical
transition is also related to the question of where the ``Heisenberg
cut'' can be applied. This notion goes back to discussions between
Werner Heisenberg and Wolfgang Pauli in 1935 and refers to the scale
of a problem in quantum theory at which a classical description can be
used without invoking a conflict with experiment. In the above example
of the brain, the decoherence timescale gives a lower bound for the
cut, guaranteeing the validity of an effective classical description
of neural processes. There may, however, be other situations in which
the Heisenberg cut lies outside the range of observational scales and
where thus quantum effects {\em can} play a role even in macroscopic
situations.

Such situations may occur when the gravitational interaction becomes
relevant. A theory of quantum gravity is not yet available in complete
form, but it seems that such a theory is needed as a major part of a
unified final theory. One reason for this belief is the incompleteness
of general relativity as expressed in the singularity theorems. \textsc{Geroch} and
\textsc{Hartle} (1986) have argued that such a theory contains
undecidable statements, at least in present formulations of the theory
that make use of path integrals. In this formulation, Geroch and
Hartle argue, no computer can carry out a computation of expectation
values, basically because the question of whether two four-dimensional
manifolds have the same topology is undecidable; in the path integral,
all possible topologies are superposed, so no calculation can be
performed. 

Another important application of G\"odel's theorems to quantum gravity
is in the canonical formulation of the theory (\textsc{Isham}
1992, \textsc{Kiefer}~2012). The quantum-gravitational wave function
$\Psi$ is there to be 
determined as a solution to the Wheeler--DeWitt equation, which is of
the form $H\Psi=0$, with $H$ being the Hamilton (energy) operator of
all degrees of freedom. In the usual formalism of quantum theory, this
equation only makes sense if the value $0$ is contained in the discrete
spectrum of $H$. But a decision about this question runs into the
spectral-gap problem discussed above: it is undecidable whether there
is indeed a gap between zero and other eigenvalues (as desired) or
not. To our knowledge, this important point has not yet been addressed
in the quantum gravity literature.

\section{Can we decide whether
  a unified physical theory is the final one?}

Kurt G\"odel, although being mathematician, had interests ranging far
beyond mathematics. In his own words:

\begin{quote}
I am apparently neither talented nor interested in 
combinatorial thinking (card games and chess, and poor memory). 
I am apparently talented and interested in conceptual thinking.
I am always interested only in how it works \ldots (and not in the
actual execution). Therefore, I should dedicate myself to the 
foundations of the sciences (and philosophy). This means: Not only the
foundations of physics, biology and mathematics, but also 
sociology, psychology, history \ldots. This
means an overview of all sciences and then foundations (which is
also what I am primarily interested in). (\textsc{Gödel} (2020),
p.~346)\footnote{German original: 
  Kombinatorisch scheine ich weder begabt noch interessiert
zu sein (Karten- und Schachspiel, und schlechtes Gedächtnis).
Begrifflich scheine ich begabt und interessiert zu sein. Es interessiert
mich bei allem nur, wie es \ldots geht (nicht die
tatsächliche Ausführung). Also soll ich mich den Grundlagen der
Wissenschaften (und der Philosophie) widmen. Das bedeutet:
Nicht nur Grundlagen der Physik, Biologie und Mathematik,
sondern auch der Soziologie, Psychologie, Geschichte \ldots.
Das heißt Überblick über sämtliche
Wissenschaften und dann Grundlagen (das ist auch, worauf ich mich
eigentlich interessiere). (\textsc{Gödel} (2020), p.~81)}
\end{quote}

Among his main other interests was physics, for which he envisaged an
underlying reality in the same sense as in mathematics. He believed
``that a question not decidable now has meaning and may be decided in
the future.''  (\textsc{G\"odel}~1990, p.~170). Can we answer the
question posed in the title of this section, whether a unified theory
is the final one or not?

It is certainly not an easy task to construct a candidate theory in
the first place. For G\"odel, the role of intuition in research was of
great importance. This he had in common with Einstein, who emphasized
the importance of intuition in his work at various places, for example
in \textsc{Einstein} (1949). Whether intuition works, is of course not
clear. Einstein was successful in constructing general relativity, but
he failed in constructing a unified field theory. As Friedrich
D\"urrenmatt remarks: ``While he arrived from the empirical by
intuition to the a priori picture, he now tried [in his attempts for a
unified field theory] to arrive by intuition from the a priori
[i.e. mathematical] description to the empirical''
(\textsc{D\"urrenmatt} 1986, p.~167).\footnote{The German original
  reads: ``Gelangte er vom Empirischen durch Intuition zum
  Apriorischen, versuchte er nun, durch Intuition vom Apriorischen zum
  Empirischen zu gelangen.''} But without connecting the
mathematical formalism to experiments or observations, all efforts may
be in vain. 

In most attempts to construct a final theory, the underlying concept
of space (or space-time) is that of a continuum. In general
relativity, space-time is modelled as a (pseudo-)Riemannian manifold,
which locally looks like $\mathbb R^4$ and thus possesses the same
cardinality as $\mathbb R$, namely $\aleph_1$. Similar features apply
to the spaces employed in other approaches, such as canonical quantum
gravity or string theory. Take the latter as an example. The theory is
defined on a ten- or eleven-dimensional manifold. To understand why
we observe only four dimensions, one must invoke a mechanism to render
the remaining dimensions unobservable. This can be achieved, for
example, by compactifying them in the form of Calabi--Yau
spaces.\footnote{The number of possible compactifications was
  estimated to be at least of the order $10^{272,000}$
  (\textsc{Douglas}~2019), which means that it is not possible to
  derive the Standard Model from string theory in any reasonable way.}
But even these spaces are manifolds and thus possess an 
uncountable number of degrees of freedom. One thus has to face the
continuum hypothesis, which we know is undecidable.

The continuum was imagined by many mathematicians to represent the
real numbers in the sense of a point set. This picture turned out
to be a very powerful one for the development of mathematics, but it
is not the only one. In the field of non-standard analysis
(\textsc{Schmieden} and \textsc{Laugwitz}~1958) developed
over the last decades, consistent models of the continuum were
constructed where many more numbers of different cardinality find
place on the continuum. Such non-standard models exist with arbitrary
high cardinality, much beyond the cardinality of the real numbers;
see, for example, \textsc{Marcja} and \textsc{Toffalori} (2003). These
numbers are also called hyperreal numbers; in such models,
a number such as $0.999\ldots$ is not equal to $1$ (as we learn in
school), but is strictly smaller than it. These non-standard models
put the notion of infinitesimals on a sound footing. One may call the
continuum as being {\em inexhaustible}, uncomparable to the picture of
a point set with the points representing real numbers. Therefore, as
long as we use the notion of a continuum at a fundamental level,
G\"odel's undecidability theorems apply and we will never know the
microstructure of space-time. The final answer about continuum or
discrete space can eventually only come from the empirical. 

The idea of mathematical realism is put to an extreme by Max Tegmark,
who entertains the idea that our Universe, in fact, consists of
mathematical structures in a realistic sense (\textsc{Tegmark~2014}). In order to avoid
problems with G\"odel's theorems, he makes the assumption that only
computable numbers are realized in Nature. But this assumption is
already in contradiction with the undecidability of the spectral-gap
problem discussed above, a problem that occurs in standard quantum
theory. Tegmark's world would be plagued with undecidability problems.     

 That there should be no infinities in our actual world was already
 emphasized by Hilbert. \textsc{Ellis} {\em et al.}, too, adopt this
 point of view and argue that ``infinity'' in physics always means potential
 infinity in the sense of very large numbers and that actual infinity
 (which they call essential infinity) does not occur. In this case,
 all antinomies and paradoxes connected with infinities
 vanish. Mathematical procedures such as regularization and
 renormalization in quantum field are then only of preliminary nature
 and would become obsolete in a final theory.

 A finite world would also lead to a finite number of superposed
 quantum states in the situations of entanglement discussed
 above. For a finite number of quantum degrees of freedom, it seems 
 that the probability interpretation of quantum theory can be derived
 without invoking a wave-function
 collapse and an {\em ad hoc}-rule.\footnote{See, for example, the discussion in
   \textsc{Kiefer} (2022), p.~94.} No problems connected with
 Heisenberg's cut would remain.

 Candidates for a unified final theory usually employ a continuous
 picture of space-time. Some notable exceptions include Carl Friedrich
 von Weizs\"acker's {\em Urtheorie} and John Wheeler's models of {\em It
   from Bit}. So far, these ideas have not led to a final theory that
 is both complete and empirically testable, but it is imaginable that
 a final theory will make use of such structures. 

 Even for a finite world, the number of degrees of freedom may be very
 large. Seth Lloyd has estimated the amount of information that the
 observable part of the Universe can register and arrived at the
 number of $10^{120}$ bits (\textsc{Lloyd} 2002). This gives an upper
 bound to the amount of possible computation, from which we shall of
 course stay far away in any practical application. Numbers with this
 order of magnitude are prevalent in cosmology and originate from the
 assumption of a smallest spatial scale of the order of the Planck
 length.\footnote{The Planck scale follows from combining Planck's
   constant, the speed of light, and the gravitational constant into a
   quantity with unit of length; it is of the order of $10^{-35}$
   metres.} It is the large size of this number and the corresponding
 smallness of the Planck scale that allow the consistent formulation
 of physical theories with an underlying space-time continuum, even if
 our ``actual'' space-time is of discrete nature.

 It thus seems that we could decide, at least in principle, whether a
 given theory is final or not only if the world were finite at small
 and large scales.\footnote{Our line of arguments is different from
   the one in \textsc{Ben-Ya'acov}~(2019), where observer
   participation in the Universe plays an essential role.}  Long ago, Bernhard
    Riemann already made speculations in this direction, although he
    was completely unaware of later developments in physics and
    mathematics. In his famous {\em habilitation thesis}, he writes:  

\begin{quote}

The question of the validity of the hypotheses of geometry in the
infinitely small is bound up with the
question of the ground of the metric relations of space.\ldots
Either therefore the reality which underlies space must form a
discrete manifoldness, or we must seek the ground of its metric
relations outside it, in binding forces which act upon
it. (\textsc{Jost} 2016, p.~40)\footnote{The German
  original reads (\textsc{Jost} 2013, p.~43):
  ``Die Frage \"uber die G\"ultigkeit der Voraussetzungen der Geometrie im
Unendlichkleinen 
h\"angt zusammen mit der Frage nach dem innern Grunde der
Massverh\"altnisse des Raumes. \ldots
Es muss also entweder das dem Raume zu Grunde liegende Wirkliche
eine discrete Mannigfaltigkeit bilden, oder der Grund der
Massverh\"altnisse ausserhalb, in darauf wirkenden bindenden
Kr\"aften, gesucht werden.'' 
The English translation is by William Clifford. The ``binding forces
which act upon it'' are in our picture affected by the undecidability theorems.}  

\end{quote}

So we conclude that, unless the space-time structure is
fundamentally discrete and the total number of degrees of freedom in
the world is finite, the question whether a given theory is the final
one or not will remain undecidable and so there will forever remain an {\em
  ignorabimus}.

\section*{Acknowlegements}

I thank Robert Helling and \L ukasz St\k epie\'n for their comments on
my manuscript.  



\end{document}